\documentclass[%
 reprint,
 amsmath,amssymb,
 aps,pra,
]{revtex4-2}
\usepackage{graphicx}
\usepackage{dcolumn}
\usepackage{bm}
\usepackage{color}

\begin{document}

\preprint{APS/123-QED}

\title{Predictions for and lack of maximal information transmission \\ in the neuromuscular junction}

\author{Eitan Goldfein}
 \affiliation{Pitzer College}
\author{Sarah Marzen}%
 \email{smarzen@natsci.claremont.edu}
\affiliation{
Department of Natural Sciences, Pitzer and Scripps College
}%
\affiliation{Kravis Department of Integrated Sciences, Claremont McKenna College}
\affiliation{National Institute for Theory and Mathematics in Biology}

\date{\today}

\begin{abstract}
A key question in theoretical biology is how effectively biological systems preserve information about their inputs while operating under physical and functional constraints. We examine that question at the neuromuscular junction (NMJ) by studying how neurotransmitter concentration is transformed into current at both cholinergic and glutamatergic NMJs. An information maximization analysis was used to derive a theoretical distribution over neurotransmitter concentrations based on biological understandings of dose-response relationships. These theoretical distributions were compared to an experimentally derived distribution obtained from a Drosophila NMJ. The theoretical and experimental distributions showed very little agreement, indicating that the glutamatergic Drosophila NMJ does not shape its distribution of synaptic vesicle release probabilities in order to maximize information transmission from nervous system to muscle. Predictions for cholinergic systems are provided.
\end{abstract}

\maketitle


\section{Introduction}

A key question in neuroscience and systems biology is how effectively biological subsystems transmit information from their inputs to their outputs. Biological signaling pathways convert physical or chemical stimuli towards cellular responses through complex molecular mechanisms, and understanding how faithfully these conversions preserve information about the input has become an important goal in theoretical biology, specifically in systems that can be framed as communication channels \cite{shannon1948mathematical,waltermann2011information,barlow1961possible}. One influential theoretical perspective proposes that biological systems may operate near the limits predicted by information theory. In this view, biological subsystems are organized to transmit as much information as possible about relevant environmental variables \cite{barlow1961possible,strong1998entropy,bell1996edges,bell1997independent,olshausen2004sparse}. If this is the case, then the input-output relationships of biological systems, and the inputs and outputs themselves, may reflect design principles consistent with optimal information transmission, such as in Refs. \cite{tkavcik2009optimizing,tkavcik2008information}.

However, many biological systems must balance information transmission against other functional constraints \cite{lan2012energy,klinger2025universal,wolpert2024stochastic}. Biological signaling networks function within constraints set by energy consumption, molecular noise, response speed, and stability to perturbations. From this perspective, systems may not maximize information transmission directly, but instead operate at compromise points where multiple competing objectives are balanced. Under this view, biological subsystems may sacrifice some potential information transmission in order to keep stability, reliability, or energetic efficiency \cite{martins2011trade,bhui2021resource,lieder2020resource}.

Alternatively, biological systems may just be ``good enough'' without any recourse to optimization or constrained optimization.

Empirical evidence supports the perspective of maximal information transmission in some circumstances, and remains ambivalent about the second and third perspective as something that seems suboptimal can be optimal with several constraints \cite{tjalma2023trade}. In some biological systems, signaling appears to approach information-theoretic limits. In early stages of embryonic development, for example, gene regulatory networks appear capable of transmitting positional information with striking precision. Studies of transcription factor gradients in the Drosophila embryo suggest that gene expression patterns can encode spatial information with efficiency close to theoretical limits \cite{tkavcik2008information}. Similarly, but not identically, efficient coding theories in sensory neuroscience have shown that neural response functions often match the statistical structure of natural stimuli, such as in Refs. \cite{laughlin1981simple,strong1998entropy,bell1996edges,bell1997independent,olshausen2004sparse}. Such theories ask sensors to change, rather than asking inputs to reshape for sensors. At the same time, other biological systems appear to deviate substantially from information-maximizing designs. Recent studies of voltage-gated potassium channels, for instance, suggest that channel parameters among evolutionary lineages do not correspond to information-optimal configurations, but instead appear tuned to sustain stable electrical dynamics as well as reliable feedback control of membrane voltage \cite{duran2023not}. Those contrasting examples suggest that biological systems exist on a spectrum between information-efficient or information-maximizing regimes and constraint-dominated regimes.

The neuromuscular junction provides a useful system for investigating these questions. At this synapse, neurotransmitters released from the presynaptic motor neuron bind to receptors on the postsynaptic membrane of the muscle cell. Binding of neurotransmitters induces conformational changes in the receptor that allow ion flow through the channel, leading to depolarization of the muscle membrane. Through this mechanism, the neuromuscular junction (NMJ) converts a chemical signal into an electrical response that ultimately results in muscle contraction \cite{bouzat2018nicotinic}.

This transformation can be understood as defining an input-output system. Here, the input is the neurotransmitter concentration sensed by receptors in the synaptic cleft, and the output is current that flows. From this perspective, the NMJ provides a useful setting for investigating how much information about neurotransmitter concentration is retained in current. Here, that question is addressed using fitted dose-response $p_{open}(c)$ curves and an information maximization analysis to study how receptor activation varies with neurotransmitter concentration and what that relationship may imply about information transmission, with methods laid out in Section \ref{sec:methods}. We test theoretical order-of-magnitude predictions against empirical relationships at the Drosophila NMJ \cite{peled2011optical} in Section \ref{sec:results} and find surprising evidence that information transmission may not be maximized at the Drosophila NMJ.

\section{Background}

\subsection{Information theory and infomax}

Information theory gives a useful framework for examining how effectively biological systems encode and transmit signals. Within this framework, a biological subsystem can be viewed as a communication channel that maps an input variable $X$ onto an output response $Y$. These two variables are random variables with realizations $x$ and $y$, respectively, and associated joint probability distribution $p(x,y)$ or joint probability density function $\rho(x,y)$. The amount of information retained through this process can be quantified using mutual information $I[X;Y]$, which measures how much uncertainty about the input is reduced once the output is observed \cite{shannon1948mathematical}. In particular, the uncertainty in the input can be quantified by the entropy in bits
\begin{equation}
H[X] = -\sum_x p(x)\log_2 p(x)
\end{equation}
and the uncertainty in the input upon knowing the output is reduced to the conditional entropy
\begin{equation}
H[X|Y] = -\sum_x \sum_y p(x,y) \log_2 p(x|y),
\end{equation}
meaning that the reduction in uncertainty of the input upon knowing the output is
\begin{eqnarray}
I[X;Y] &=& H[X]- H[X|Y] \\
&=& \sum_{x,y} p(x,y) \log_2 \frac{p(x|y)}{p(x)}.
\end{eqnarray}
This expression can be rewritten into many forms using rules for conditional probabilities, such as $p(x,y)=p(x)p(y|x)=p(y)p(x|y)$. Some straightforward rewriting will show that the mutual information is symmetric in $X$ and $Y$, and that in fact, it is also the reduction in uncertainty about output upon knowing input. If one has probability density functions because $X$ and $Y$ are continuous random variables, one simply replaces probability distributions with probability density functions and sums with integrals:
\begin{eqnarray}
I[X;Y] &=& \int_{x,y} \rho(x,y) \log_2 \frac{\rho(x,y)}{\rho(x)\rho(y)} dx dy.
\end{eqnarray}
The expression for mutual information is well-defined in that if continuous random variables are approximated as discrete random variables, the answer is basically the same, assuming you discretize finely enough. Entropy is not \cite{cover1999elements}.

The function for entropy may seem arbitrary, but it actually is the only function up to a positive multiplicative constant (a change in basis of the log) of probability distributions that satisfies three reasonable desiderata \cite{shannon1948mathematical,cover1999elements}. Thus, the mutual information acquires a number of interpretations. One interpretation is that it is the reduction in uncertainty about input if output is known, as previously stated, or the reverse. Another interpretation is that it is also a measure of correlation between input and output \cite{kinney2014equitability} that can capture both linear and nonlinear correlations because it quantifies the Kullback-Liebler divergence between the joint distribution and the product of the marginals, and so measures deviation from complete independence. Finally, and of particular interest to us, its supremum over input with fixed information channel $p(y|x)$ has an operational definition as the maximum amount of information that can be passed through the information channel \cite{shannon1948mathematical,cover1999elements}.

We ignore in this paper the very interesting proposition that rate-distortion theory is relevant to the functioning of organisms. It may. But in this paper, the organism controls not the information channel (which is set by the biophysical channel) but instead controls the input distribution $\rho(x)$. Rate-distortion theory applies when the information channel $\rho(y|x)$ is under the organism's control. Here, that information channel is set by the unavoidable biophysical variability in receptor openings and closings. For a listing of how rate-distortion theory may be relevant to organism functioning, please see Refs. \cite{marzen2017evolution,marzen2025resource} and references therein.

A fundamental question we can ask as biophysicists and as interdisciplinary scientists is whether or not biology cares about bits \cite{bialek2012biophysics,barlow1961possible}. Does biology measure anything in bits implicitly? Many of the major theorems that frequent the information theory literature have strict assumptions that do not apply to the biological problems we often face \cite{cover1999elements}. Even so, there has been some surprising success using information theory to understand biological systems \cite{barlow1961possible,strong1998entropy,bell1996edges,bell1997independent,olshausen2004sparse}, likely meaning that at least one interpretation of mutual information is relevant to biological organisms.

Over the past two decades, this perspective has been applied to a broad range of biological systems, including gene regulation, sensory processing, and cellular signaling networks \cite{shannon1948mathematical,tostevin2009mutual,waltermann2011information}. An important example comes from early embryonic development, where Ref. \cite{tkavcik2008information} studied the relationship between transcription factor concentration and gene expression and found that the system appeared to convey positional information with efficiency close to its theoretical maximum. Taken together, these studies suggest that information theory yields a quantitative way to connect biological mechanisms with the performance of signal transmission.

Information-theoretic ideas have also been applied more directly to biochemical communication, where cells must distinguish real changes in input from noise. In that setting, mutual information provides a way to ask how reliably an input can be inferred from a downstream response, even when the signaling process is noisy and not completely linear. This makes the framework especially useful for biological systems in which the input-output relationship is probabilistic rather than exact \cite{waltermann2011information,cheong2011information}, as was true for the aforementioned application in Ref. \cite{tkavcik2008information} but that was not true for applications in Refs. \cite{laughlin1981simple,bell1996edges}.

Information-theoretic approaches have contributed to the idea that some biological systems may be organized to preserve as much relevant information as possible about their inputs. From this perspective, response properties should reflect an efficient use of available dynamic range and sensitivity in order to generate informative outputs \cite{barlow1961possible}. At the same time, biological signaling systems are not governed by information transmission alone. They must also operate under constraints associated with noise, response speed, robustness, and energetic cost \cite{lan2012energy}. Consequently, such systems may not directly maximize mutual information, but instead occupy compromise points determined by multiple competing demands. Studies of receptor-based sensing systems have shown that properties such as dynamic range, gain, noise, and response time cannot be optimized independently, because improvement in one feature may come at the expense of another \cite{martins2011trade}. This broader perspective is important for the present thesis, since it suggests that even if receptors retain meaningful information about ligand concentration, this does not necessarily mean that they are organized solely to maximize information transmission \cite{duran2023not}.

\subsection{The neuromuscular junction (NMJ) as an input-output system}

The neuromuscular junction is the synapse through which motor neurons communicate with muscle fibers. When the NMJ is cholinergic, when an action potential reaches the presynaptic terminal, acetylcholine is released into the synaptic cleft, where it diffuses across the cleft and binds to nicotinic acetylcholine receptors (nAChRs) on the postsynaptic membrane of the muscle cell. For glutamatergic systems, action potentials lead to glutamate binding to ionotropic glutamate receptors (iGluRs) on the postsynaptic membrane of the muscle cell. Binding of neurotransmitters induces conformational changes in the receptor that allow ions to pass through the channel, producing depolarization of the muscle membrane and ultimately playing a role in the electrical response that leads to muscle contraction. Because this process links a chemical input to a measurable postsynaptic response, the neuromuscular junction provides a useful system for studying how biological signaling preserves information about its input. In the context of this thesis, the relevant input variable is acetylcholine or glutamate concentration, while the output is receptor activation, summarized by the probability that a receptor is in the open state \cite{bouzat2018nicotinic,unwin2013nicotinic,karlin2002emerging}.

Synaptic transmission is also not uniform across release sites. Prior work has shown that release probability can vary substantially from one site to another, including at the Drosophila neuromuscular junction, which is glutamatergic unlike vertebrates. That matters here because it means synaptic output is formed not just by average release, but also by how probability is distributed across individual sites. In other words, variability across sites is part of the biology, not just noise around a single mean value \cite{atwood2002diversification,peled2011optical}.

This framing does not mean that the postsynaptic muscle is trying to preserve every random fluctuation in vesicle release. Rather, variable release is treated as one way that the acetylcholine input can vary before it reaches the receptor population. The receptors then add their own noise, since individual channels open and close probabilistically. The question in this thesis is therefore not whether the muscle directly encodes vesicle-release noise, but how much information about the local chemical input can be preserved through receptor activation.

\section{Methods}
\label{sec:methods}

In general, it is quite difficult to calculate optimal input distributions for maximizing information transmission for a channel in which $x$ is input and $y$ is output, even if the input and output are scalars, as we consider in this manuscript. To be clear, we are searching for the optimal input distribution $\rho^*(x)$ that maximizes:
\begin{eqnarray}
\rho^*(x) &=& \arg\max_{\rho(x)} I[X;Y] \\
&=& \arg\max_{\rho(x)} \int\int \rho(x) \rho(y|x) \log_2 \frac{\rho(y|x)}{\rho(y)} dx dy
\end{eqnarray}
However, if the communication channel is approximately conditionally Gaussian so that
\begin{equation}
\rho(y|x) \approx \frac{1}{\sqrt{2\pi \sigma_{y|x}^2}}e^{-(y-\bar{y}(x))/2\sigma_{y|x}^2}
\end{equation}
where the noise $\sigma_{y|x}$ is small, then it has been shown through variational calculus \cite{tkavcik2009optimizing} that the optimal input distribution is
\begin{equation}
\rho^*(x) = \frac{1}{Z}\frac{|\frac{d\bar{y}(x)}{dx}|}{\sigma_{y|x}}.
\end{equation}
Here, $Z$ is a normalization constant:
\begin{equation}
1 = \int \frac{|\frac{d\bar{y}_{|x}}{dx}|}{\sigma_{y|x}} dx.
\end{equation}
From this optimal distribution over inputs comes an optimal distribution over outputs:
\begin{eqnarray}
\rho^*(y) &=& \int \rho(y|x) \rho^*(x) dx.
\end{eqnarray}
While the distribution over outputs has a complicated form, it mirrors that of Fig. \ref{fig:hernan} in the cases shown here and peaks at low and high currents. The distribution over $\bar{N}_{open}$, on the other hand, is quite simply proportional to $1/\sigma_{y|x}$, which for the channels considered here again peaks at low and high currents.
The channel capacity becomes
\begin{equation}
C = \log_2 \frac{Z}{\sqrt{2\pi e}}.
\label{eq:capacity1}
\end{equation}
For a detailed and accessible derivation, see Ref. \cite{marzen2013statistical}.
Rather than use the Blahut-Arimoto equation that applies to general information channels, we will use simply this small-noise approximation.

\section{Results}
\label{sec:results}

\begin{figure}
\includegraphics[width=0.5\textwidth]{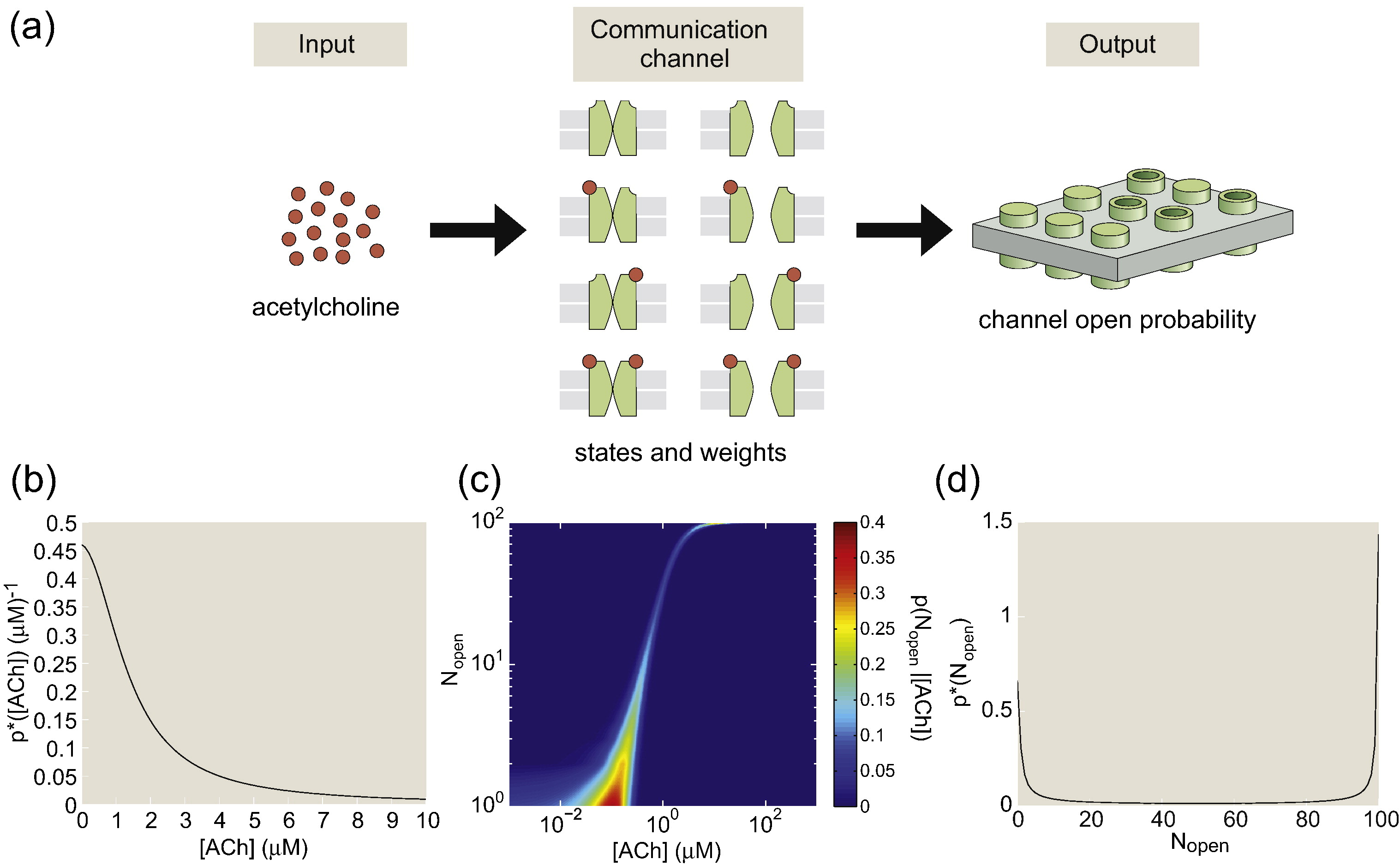}
\caption{At top, the biophysical channel in which neurotransmitter concentration causes ion channels to open and close, leading to current flowing at a cholinergic NMJ into the muscle fiber. At bottom, the information channel in which an optimal input distribution over neurotransmitter concentrations given the biophysical channel shown in the center yields an optimal output distribution over the number of open receptors. At bottom left, the optimal input distribution. At bottom middle, the biophysical channel that communicates ligand concentration to open probabilities (currents). As described in later sections, the information channel assumes a binomial distribution for the number of open receptors with probabilities given by physiological dose-response relationships. At bottom right, the optimal output distribution over the number of open channels.}
\label{fig:hernan}
\end{figure}

With this results section, we aim to make it straightforward for any practitioner to implement similar calculations if more details are known about how current flows into muscles via ion channels opening and closing. The basic idea behind these calculations is contained simply in Fig. \ref{fig:hernan}, reproduced with permission from Ref. \cite{marzen2013statistical}. The neuromuscular junction takes in information about what the nervous system wants to do via neurotransmitter concentrations and converts it into muscle activity via current flow. Thus, the neurotransmitter concentrations flow through an information channel that converts concentrations (commands from the nervous system) into current (muscle activity). The biophysical implementation of this channel is, at its simplest, neurotransmitters acting as ligands for either nicotinic acetylcholine receptors for cholinergic NMJs or ionotropic glutamate receptor for glutamatergic NMJs. When neurotransmitters are present, the probability of the ion channel/receptor opening and letting current flow is larger. The degree to which the current flow increases determines the information channel and thus the optimal distribution of inputs-- the probability density function over neurotransmitter concentrations. This can be empirically measured in tour de force experiments such as Ref. \cite{peled2011optical} indirectly by measuring the probability distribution of synaptic vesicle release probabilities, as each vesicle contains a quantum (set small amount) of neurotransmitters and therefore increases the neurotransmitter concentration in a predictable fashion.

In what follows, we aim for an order of magnitude calculation of a complicated information channel. One major assumption we make is that the voltage is roughly kept constant by the opening and closing of other ion channels that do not respond to the neurotransmitter. This assumption is biologically reasonable, and it simplifies the calculation greatly, because then the collective activity of a number of receptors is described by a binomial distribution instead of something much more complicated. If a finer calculation is desired, one would use the cable equation to model the membrane voltage change as current flows. If we acknowledge that voltage also affects the probability of ion channels opening, this cable equation leads to a mean-field voltage coupling between otherwise independent receptors. If membrane voltage is instead roughly constant, open probabilities are depressed by a constant value that still allows for independence of the various ion channels, so that their collective activity is still described by a binomial distribution. Then, the dose-response curves measured at physiological conditions completely describe the open probability of the ion channels. If equilibrium membrane potential is altered, the optimal input distribution changes in a way that is described in Appendix \ref{app:1}.

\subsection{Optimal input distributions for neurotransmitter concentrations}

We start by identifying optimal input probability density functions over ligand concentration for cholinergic and glutamatergic NMJs. These optimal input probability density functions, which we call optimal input distributions for shorthand, are based on a construction of an information channel from the biophysical channel. The idea is rather simple to state-- we have some number of ion channels that respond to neurotransmitters by letting current through in a way that depends on the concentration of neurotransmitters. If we can characterize quantitatively exactly how this works, we can then manage to find the information channel from biophysical principles. For order-of-magnitude results, we use the phenomenological Hill model to describe dose-response relationships, as these fit the empirical data quite well and are readily available \cite{prince1999acetylcholine,han2015functional}. More detailed analyses can use statistical mechanical models \cite{phillips2012physical,marzen2013statistical} to find dose-response relationships. This matters when the dynamic range assumed by the Hill model (which is $1$) is obviously larger than the real dynamic range, where dynamic range is defined as the maximum open probability minus the minimum open probability. For the ion channels assessed here, the Hill model was adequate for order-of-magnitude calculations, as the dynamic range of both nAChRs and iGluRs is quite large.

\subsubsection{The approximate cholinergic information channel}

For cholinergic systems, acetylcholine floods the synapse through release of synaptic vesicles, and this acetylcholine attaches to nAChRs. From Ref. \cite{prince1999acetylcholine}, their dose-response relationship when modeled phenomenologically by a Hill model is
\begin{eqnarray}
p_{open,nAChR} &=& \frac{c^{n_H}}{K_d^{n_H}+c^{n_H}}
\label{eq:Hill}
\end{eqnarray}
with adult nAChRs having a $K_d=1.6\mu M$ and $n_H=1.3$ and fetal nAChRs having a $K_d=1.3\mu M$ and $n_H=1.4$.
If we assume that the amount of current is simply proportional by a constant $J_0$ to the number of open nAChRs $N_{open}$, then we immediately have the information channel, or $p(J|c)$ where $J$ is the current. The current $J$ is proportional to $N_{open}$, which acts as a binomial distribution:
\begin{eqnarray}
p(N_{open}|c) &=& {N \choose N_{open}} p_{open}(c)^{N_{open}} (1-p_{open}(c))^{N-N_{open}} \\
p(J|c) &=& p(N_{open}=\frac{J}{J_{0}}|c) |\frac{dN_{open}}{dJ}| \\
&=& \frac{1}{J_0} p(N_{open}=\frac{J}{J_{0}}|c).
\end{eqnarray}
In these expressions, $N$ is the total number of nAChRs.

Although this expression looks complicated, there are always so many receptors that the small-noise approximation from Sec. \ref{sec:methods} applies. In the large $N$ limit, the Central Limit Theorem applies, and the binomial distribution becomes approximately Gaussian to a very good approximation. Furthermore, as the number of receptors $N$ increases, the ratio of the standard deviation of the current to the mean current decreases as $\frac{1}{\sqrt{N}}$. This means that we can use the small-noise approximation from Sec. \ref{sec:methods} with mean current
\begin{eqnarray}
\bar{J} &=& J_0 N p_{open,nAChR}
\end{eqnarray}
and variance of the current
\begin{eqnarray}
\sigma_J^2 &=& J_0^2 N p_{open,nAChR}(1-p_{open,nAChR}).
\end{eqnarray}
Thus, we have characterized the cholinergic information channel at the NMJ as roughly Gaussian with a mean and variance that depend completely on physiological dose-response relationships.

This then allows us, using the small-noise approximation, to directly find that
\begin{eqnarray}
\rho^*(c) &=& \frac{1}{Z}\frac{|\frac{d\bar{J}}{dc}|}{\sigma_J} \\
&=& \frac{\sqrt{N}}{Z} \frac{|\frac{dp_{open}}{dc}|}{\sqrt{p_{open}(1-p_{open})}}.
\end{eqnarray}
Note that the important biophysical constant $J_0$ has no real effect on the information channel, as it is simply a proportionality factor that drops out of the analysis. Similarly, the constant $\sqrt{N}$ is used as part of the normalization factor rather than affecting the shape of the distribution. The normalization constant $Z$ is simply obtained from
\begin{eqnarray}
1 &=& \int_0^{\infty} \rho^*(c) dc
\end{eqnarray}
which implies
\begin{eqnarray}
Z &=& \sqrt{N}\int_{p_{open}^{(min)}}^{p_{open}^{(max)}} \frac{dp}{\sqrt{p(1-p)}} \\
&=& 2\sqrt{N} \left(\sin^{-1}\sqrt{p_{open}^{(max)}}-\sin^{-1}\sqrt{p_{open}^{(min)}}\right).
\end{eqnarray}
This implies that that the optimal distribution over neurotransmitter concentrations is
\begin{equation}
\rho^*(c) = \frac{1}{2\left(\sin^{-1}\sqrt{p_{open}^{(max)}}-\sin^{-1}\sqrt{p_{open}^{(min)}}\right)}\frac{|\frac{dp_{open}}{dc}|}{\sqrt{p_{open}(1-p_{open})}}.
\end{equation}
In other words, with the approximation that membrane voltage is constant, the only thing that determines the optimal distribution over neurotransmitter concentrations is the physiological dose-response relationship $p_{open}(c)$.

\begin{figure}
\includegraphics[width=0.5\textwidth]{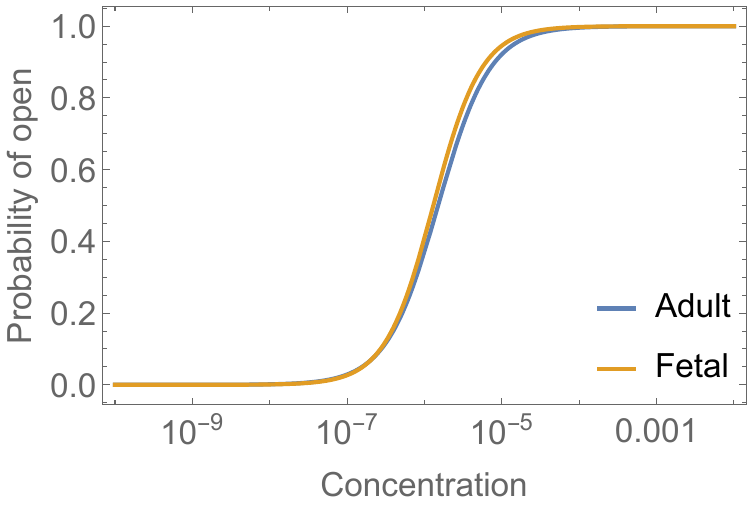}
\includegraphics[width=0.5\textwidth]{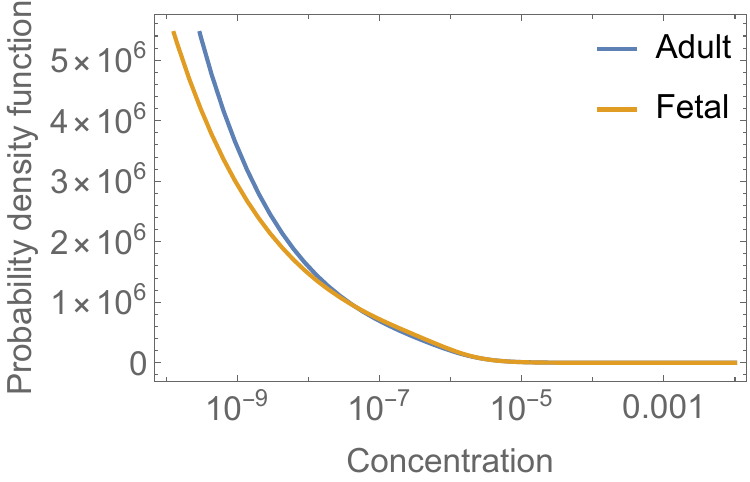}
\caption{At top, we show the phenomenological fits to dose-response data in physiological conditions for nAChRs that are from fetal mice and adult mice with concentration in molar. The phenomenological fits are simply Hill functions with parameters taken from Ref. \cite{prince1999acetylcholine} and shown in Eq. \ref{eq:Hill} with adult nAChRs having a $K_d=1.6\mu M$ and $n_H=1.3$ and fetal nAChRs having a $K_d=1.3\mu M$ and $n_H=1.4$. At bottom, we show the optimal input distribution over neurotransmitter concentration that results with concentration in molar. Small changes in dose-response relationships result in measurable changes to the optimal input distribution.}
\label{fig:optimalinputcholinergic}
\end{figure}

In Fig. \ref{fig:optimalinputcholinergic}, we can see that the optimal distribution over neurotransmitter concentrations is somewhat sensitive to details of the dose-response relationship. Two slightly different dose-response relationships, one for fetal nAChRs and one for adult nAChRs, can lead to slight but noticeable differences between optimal input distributions over neurotransmitter concentration. Predicting these differences-- which surprisingly occur in the regime where the open probability is roughly zero-- can allow for targeted experiments that examine empirical input distributions of neurotransmitter concentration in those otherwise unstudied regimes. This means that in order to understand information transmission, dose-response relationships at saturation as open or closed must be investigated as thoroughly as possible, even though this is a regime in which small amounts of experimental error could have large consequences for an appropriate fit between statistical mechanical model and empirical data.

\subsubsection{The approximate glutamatergic information channel}

\begin{figure}
\centering
\includegraphics[width=0.5\textwidth]{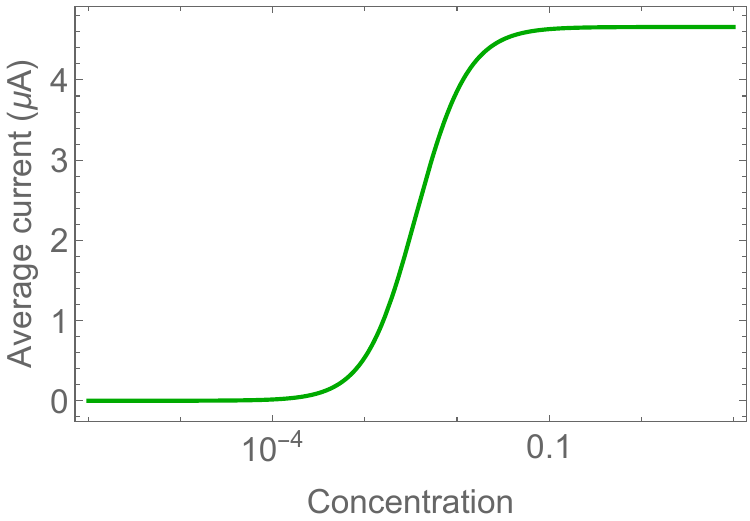}
\includegraphics[width=0.5\textwidth]{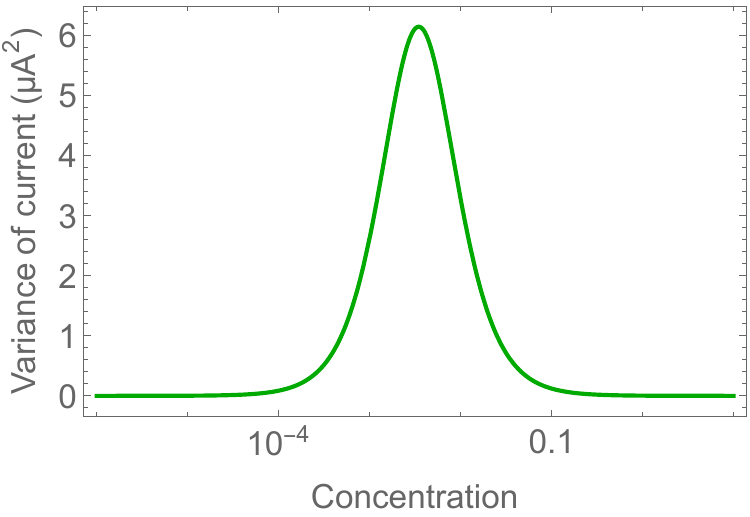}
\includegraphics[width=0.5\textwidth]{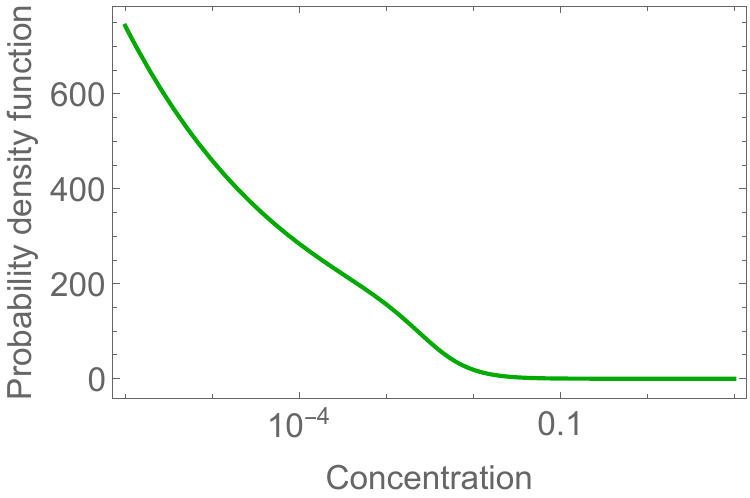}
\caption{(Top) Normalized average current $\bar{J}/N$ in $\mu A$ for the glutamatergic NMJ versus concentration of glutamate in molar. (Middle) Variance in normalized average current $\sigma_J/N$ for glutamatergic system versus concentration of glutamate in molar. (Bottom) The optimal pdf over glutamate concentration that maximizes information transmission at the glutamatergic NMJ for concentrations measured in molar. All empirical constants were taken from Ref. \cite{han2015functional}.}
\label{fig:optimalinputglutamatergic}
\end{figure}

The glutamatergic information channel is slightly more complicated than the cholinergic information channel, simply because there are two slightly different iGluRs that both play an important role in determining the amount of current that flows through the muscle fiber. Let $c$ be the concentration of glutamate. The two receptors we focus on are type A and type B receptors, which have been measured to have slightly different dissociation constants $K_d$, slightly different effective Hill coefficients $n_H$, and slightly different currents that flow at saturation $j_{A/B}$. Using again the Hill functional form, we find using empirically measured parameters from Ref. \cite{han2015functional} that
\begin{eqnarray}
p_{open,A} &=& \frac{c^{1.6}}{c^{1.6}+(3.4 mM)^{1.6}} \\
p_{open,B} &=& \frac{c^{1.5}}{c^{1.5}+(5.9 mM)^{1.5}}.
\end{eqnarray}
What we care about-- the output of the information channel-- is the current. This time, because there are two types of receptors that we are considering, the current is more complicated:
\begin{eqnarray}
J &=& j_A N_{open,A} + j_B N_{open,B}.
\end{eqnarray}
Both $N_{open,A}$ and $N_{open,B}$ are separately binomial, so in the limit of a large number of receptors of each, we find that these two distributions are roughly Gaussian. The sum of two Gaussians is Gaussian, and so as long as the number of each type of iGluR is larger than about $10$, we can approximate the current $J$ as Gaussian. Let us label $\alpha$ as the fraction of type A to type B receptors, which is roughly $\alpha=0.7$. The mean of the current is
\begin{equation}
\bar{J} = j_A \alpha N p_{open,A} + j_B (1-\alpha) N p_{open,B}
\end{equation}
and its variance is
\begin{equation}
\sigma_J^2 = j_A^2 \alpha N p_{open,A} (1-p_{open,A}) + j_B^2 (1-\alpha) N p_{open,B}(1-p_{open,B}).
\end{equation}
\begin{widetext}
Because there are two receptor types that are governing the information channel, the optimal distribution over glutamate concentration is more complicated, but still follows
\begin{eqnarray}
\rho^*(c) &=& \frac{1}{Z} \frac{|\frac{d\bar{J}}{dc}|}{\sigma_J} \\
&=& \frac{\sqrt{N}}{Z} \frac{|\frac{d}{dc}(\alpha p_{open,A} + \frac{j_B}{j_A} (1-\alpha) p_{open,B})|}{\sqrt{\alpha p_{open,A} (1-p_{open,A}) + \frac{j_B^2}{j_A^2} (1-\alpha) p_{open,B}(1-p_{open,B})}}.
\end{eqnarray}
Unlike before, there is no analytic expression for the normalization constant $Z$, but it is proportional to $\sqrt{N}$:
\begin{equation}
Z = \sqrt{N} \int_0^{\infty} \frac{|\frac{d}{dc}(\alpha p_{open,A} + \frac{j_B}{j_A} (1-\alpha) p_{open,B})|}{\sqrt{\alpha p_{open,A} (1-p_{open,A}) + \frac{j_B^2}{j_A^2} (1-\alpha) p_{open,B}(1-p_{open,B})}} dc.
\end{equation}
\end{widetext}
This time, the ratio of $j_A$ to $j_B$ is not only an important biophysical constant, but determines the information channel as well. From Ref. \cite{han2015functional}, we use $j_A = 5.8 \mu A$ and $j_B = 2.0\mu A$. The standard errors on these two values are rather large compared to the standard errors on the dissociation constants and effective Hill coefficients, but we are merely looking for order-of-magnitude estimates in this manuscript. These biophysical estimates lead to the information channel and optimal input distribution shown in Fig. \ref{fig:optimalinputglutamatergic}. As we will see, these order-of-magnitude estimates are enough to understand whether or not the Drosophila NMJ, which is glutamatergic, is maximizing information transmission. Note that having two receptors with differing dissociation constants and effective Hill coefficients means that the optimal input distribution over neurotransmitter concentration has a severely different shape than it would if there were just one receptor by comparison of Figs. \ref{fig:optimalinputcholinergic} and \ref{fig:optimalinputglutamatergic}, even though the two receptors have similar dissociation constants and effective Hill coefficients.

\subsection{Approximate channel capacity of the neuromuscular junction}

The optimal input distribution over neurotransmitter concentrations is, as we have seen from Figs. \ref{fig:optimalinputcholinergic} and \ref{fig:optimalinputglutamatergic}, very sensitive to exactly what dose-response relationships are used. The channel capacity, as it turns out, is not.

\begin{figure}
\includegraphics[width=0.5\textwidth]{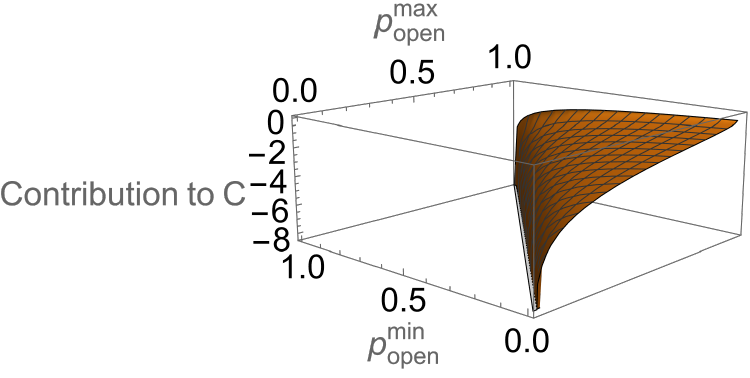}
\caption{For the cholinergic system, this shows how the contribution to channel capacity from dose-response relationships rather than from the total number of receptors, or $\log_2 \frac{z}{\sqrt{2\pi e}}$, depends on the dynamic range. As $p_{open}^{(max)}$ increases its difference from $p_{open}^{(min)}$, the channel capacity contribution greatly increases until saturation at high dynamic ranges.}
\label{fig:dynrange}
\end{figure}

\begin{widetext}
To get the channel capacity, we simply use Equation \ref{eq:capacity1} and find that
\begin{equation}
C = \frac{1}{2}\log_2 N + \log_2 \frac{z}{\sqrt{2\pi e}}
\label{eq:capacity2}
\end{equation}
with $z=\frac{Z}{\sqrt{N}}$
where for cholinergic systems,
\begin{equation}
z = 2 \left(\sin^{-1}\sqrt{p_{open}^{(max)}}-\sin^{-1}\sqrt{p_{open}^{(min)}}\right),
\label{eq:z1}
\end{equation}
and for glutamatergic systems,
\begin{equation}
z = \int_0^{\infty} \frac{|\frac{d}{dc}(\alpha p_{open,A} + \frac{j_B}{j_A} (1-\alpha) p_{open,B})|}{\sqrt{\alpha p_{open,A} (1-p_{open,A}) + \frac{j_B^2}{j_A^2} (1-\alpha) p_{open,B}(1-p_{open,B})}} dc.
\end{equation}
\end{widetext}
Note that for the cholinergic NMJ, it is clear from the analytic expression that channel capacity and dynamic range are strongly correlated. If $p_{open}^{(max)}$ increases, then so does $C$; if $p_{open}^{(min)}$ decreases, then so does $C$, which we can see just by looking at Eq. \ref{eq:z1}. When dynamic range is near $1$, this dependence is not so strong, but when dynamic range is not as high, a pressure to increase information transmission will increase dynamic range drastically. See Fig. \ref{fig:dynrange}.

\begin{figure}
\includegraphics[width=0.5\textwidth]{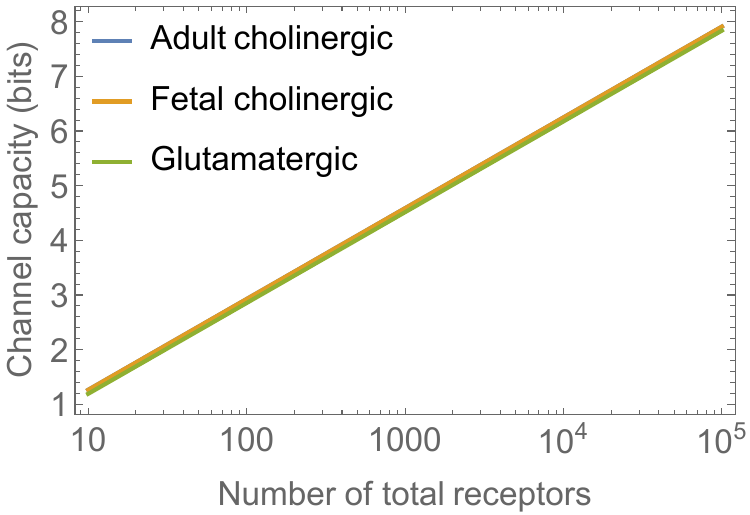}
\caption{As the total number of receptors increases, the channel capacity increases logarithmically. At biologically reasonable values of about $10,000$ total receptors, we find across NMJs of different types that the channel capacity is around $6$ bits. This channel capacity is not so sensitive to details of the biophysical channel, unlike optimal input distributions.}
\label{fig:channelcapacity}
\end{figure}

Note also that, as shown in Fig. \ref{fig:channelcapacity} and as illustrated by Eq. \ref{eq:capacity2}, the channel capacity is primarily determined by the number of receptors rather than by biophysical parameters. The biophysical parameters only affect the offset, which essentially captures the size of the dynamic range through its dependence on $p_{open}^{(max)}$ and $p_{open}^{(min)}$, while the channel capacity grows logarithmically with the number of neurotransmitters.

As dynamic range grows to $1$, the contribution to the channel capacity from biophysical parameters saturates. Because of this saturation, an order-of-magnitude calculation of channel capacity need not require complicated statistical mechanical models; oftentimes, a phenomenological model will work just as well. See Appendix \ref{app:validity}.

The parameter $N$, the total number of receptors that are either nAChRs for the cholinergic NMJ or iGluRs for the glutamatergic NMJ, is a biophysical constant that is the major governing factor in deciding channel capacity $C$. If $N$ increases, then $C$ increases logarithmically. The deviations in $z$ from the cholinergic NMJ to the glutamatergic NMJ actually have very little effect on the channel capacity $C$, as shown in Fig. \ref{fig:channelcapacity}-- the three systems (fetal cholinergic NMJ, adult cholinergic NMJ, and glutamatergic NMJ) nearly have the exact same relationship between channel capacity $C$ and number of total receptors. For order-of-magnitude estimates that place the receptor count at around $10,000$, we have channel capacities that are roughly $6$ bits. This is far higher than the channel capacities that have been measured so far in other systems. The Drosophila gene expression patterning to determine cell position \cite{tkavcik2008information} has a channel capacity of about $1.5$ bits; evolutionary channels \cite{soriano2023well} have about $2$ bits per allele; noisy biochemical signaling systems have been measured to have around $1$ bit of channel capacity \cite{cheong2011information,cohen2009dynamics,bao2010variability}; and voltage-gated potassium channels \cite{duran2023not} and bacterial chemotactic receptors \cite{cui2026apparent} have less than $1$ bit by means of a binary random variable (active or not active) being their output. If trajectories are the output of an information channel, then suddenly we can see channel capacities that are much higher, at about $2$ bits per second \cite{tostevin2009mutual}.

\subsection{Comparing to an empirical distribution at the larval Drosophila NMJ}

The key question we have to ask, constantly, is whether or not biology cares about information flow. As we have seen, the hypothesis that information flow is maximized creates testable predictions for the input distribution of neurotransmitter concentrations in the case of the NMJ. If the biological system can control these input distributions, then there is a chance that channel capacity is reached. In the case of the NMJ, a biological system could plausibly have evolved to control the variable probability of synaptic vesicle release, which would then control the probability distribution over neurotransmitter concentration.

Such measurements are rare, but in a seminal experiment many years ago, the probability distribution of the probability of synaptic vesicle release was measured. A quick order-of-magnitude calculation can reveal whether or not there is experimental evidence that information transmission is maximized at the glutamatergic Drosophila NMJ. The web program WebPlotDigitizer was used to extract the data from Fig. 5b in Ref. \cite{peled2011optical}, and these histograms were compared to the histograms that would be expected if information transmission were maximized at the Drosophila NMJ.

To compare, we must find order-of-magnitude procedures to compare apples to apples. Release probability was first converted into an expected number of released vesicles by multiplying by $50$ release sites, based on the number of release sites reported by Ref. \cite{peled2011optical}. This value was then multiplied by $10,000$ acetylcholine molecules per vesicle, based on standard estimates of acetylcholine vesicle content, and converted to moles using Avogadro’s number. Finally, this amount was divided by an estimated synaptic cleft volume. The cleft volume was estimated geometrically as an effective synaptic contact area multiplied by cleft width. An effective area of $139 \mu m^2$ was used as an approximate NMJ-scale contact area, guided by morphometric studies showing that NMJ area varies across synapse populations and can be quantified from pre- and postsynaptic morphology \cite{jones2016nmj}. The cleft width was taken to be $20 nm$, consistent with reported Drosophila NMJ cleft widths \cite{prokop1996presynaptic,koper2012analysis}. This gave $V_{cleft} = 139 \mu m^2\times 20 nm = 2.78 \times 10^{-15} L$, which gives the conversion
\begin{equation}
c = \frac{p_{release}\times 50 \times 10000}{(2.78 \times 10^{-15})(6.022\times 10^{23})}
\label{eq:conversion}
\end{equation}
in molar. Finally, probability density functions from theoretical considerations in previous sections that derived optimal input distributions can be compared to the empirical histograms using $P^*(c\in [c_0,c_1]) = \int_{c_0}^{c_1}\rho^*(c) dc$.

\begin{figure}
\includegraphics[width=0.5\textwidth]{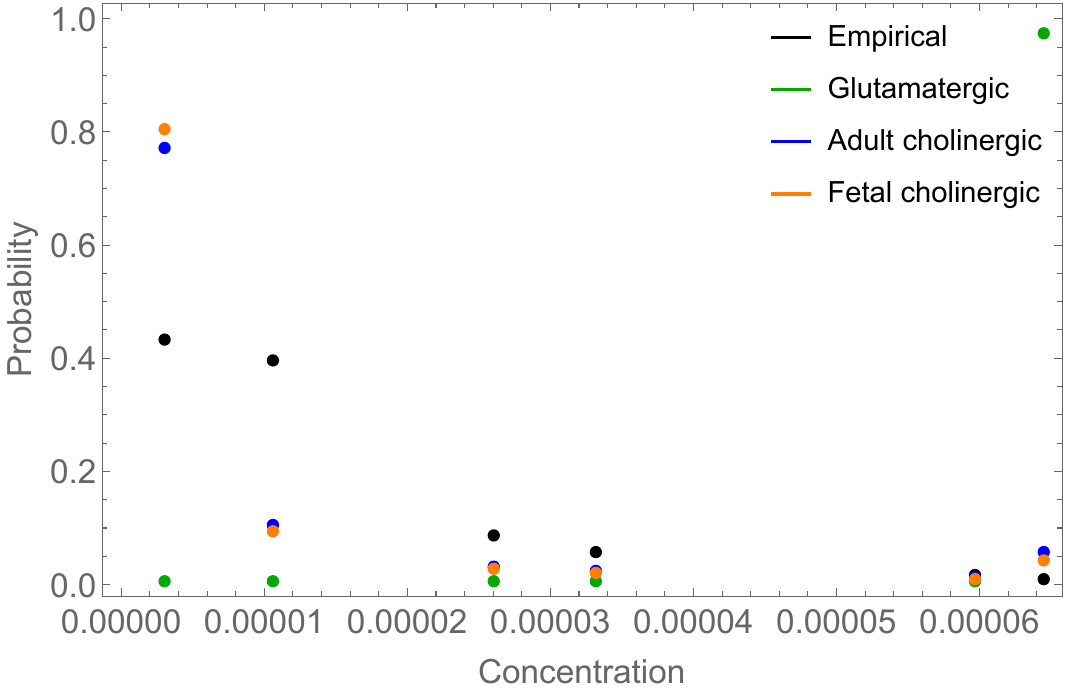}
\caption{The empirical probability distribution of synaptic vesicle release probability translates to a rough estimate of empirical histogram for concentrations of neurotransmitters using Eqs. \ref{eq:conversion} and data from Ref. \cite{peled2011optical} with concentrations in molar. The empirical probabilities do not match order-of-magnitude theoretical predictions based on maximal information transmission.}
\label{fig:comparison}
\end{figure}

The comparison between empirical histograms and the histograms that would be expected from theoretical considerations if information transmission was maximized at the Drosophila NMJ is shown in Fig. \ref{fig:comparison}. A linear scale is used for better alignment with Ref. \cite{peled2011optical}. Note that what is shown is a histogram in which each dot integrates over a complex probability density function for concentration ranges determined by the size of bars in Fig 5b of Ref. \cite{peled2011optical}. Just for comparison, the calculation for a cholinergic NMJ is shown as well. We see that there is a failure of the order-of-magnitude calculation to match the empirical measurements by such a degree that actually, it may be the case that information transmission is not maximized at the Drosophila NMJ. There are certainly fudge factors in these estimates, e.g. the number of neurotransmitters per vesicle release could range from $5,000$ to $10,000$ plausibly. But playing with these numbers does little to change the fact that the dissociation constants for glutamatergic NMJs are so different than what they would need to be for an empirical match. As a result, theoretical predictions for a cholinergic NMJ are actually much closer to what would be expected for an empirical match because the dissociation constants of nAChRs are orders of magnitude different than for iGluRs.

It would be interesting to also compare the empirical mutual information to the channel capacity to assess if the empirical probability density function over ligand concentrations allows for maximal information transmission to be approximately reached. This is unlikely, as the empirical probability density function seems to be tuned for a dissociation constant which is roughly three orders-of-magnitude smaller than the empirical dissociation constant. But more than that, without finer resolution on the histogram so as to enable us to estimate a reasonably accurate probability density function, no empirical mutual information can be obtained.

\section{Conclusion}

The results shown here explicate how one might perform information-maximizing predictions for optimal neurotransmitter concentration distributions at the NMJ. Theoretical results on minimal models showcase that dynamic range and channel capacity are highly correlated at cholinergic NMJs. Small deviations in dose-response curves can result in noticeable changes in optimal input distributions while having little effect on channel capacity estimates. The minimal models used here are easily extended to more receptor types and even voltage coupling, but what we have here suffices for an order-of-magnitude estimate of optimal input distributions and channel capacity estimates.

We would like to stress that these are order-of-magnitude estimates, not the final answer. For instance, the conversion from release probability to estimated neurotransmitter concentration also adds uncertainty. This step required assumptions about the number of release sites, the number of acetylcholine molecules per vesicle, and the effective synaptic cleft volume. The cleft volume was estimated from an approximate contact area and cleft width, rather than measured directly in the same preparation. Changing any of these values would shift the estimated concentration axis and could change the visual agreement between the theoretical and experimental distributions. Similarly, as changes in biophysical parameters like the total number of receptors can greatly change the channel capacity, the channel capacity value should be interpreted carefully. The estimate of about $6$ bits depends on the fitted response curve, the assumed receptor population size, and the binomial noise model. It is therefore an order-of-magnitude estimate under the assumptions of the model, not a directly measured biological value.

Even with the roughness of our approximations, we found evidence that surprisingly, the Drosophila NMJ may not maximize information transmission. In particular, the dissociation constants for the Drosophila NMJ iGluRs differ so greatly from what they would need to be for a match that no amount of uncertainty in quantities like the number of neurotransmitters per vesicle are going to change that the empirical histogram has most of its weight in low concentrations while the information-maximizing distribution has its weight in relatively high concentrations.

However, we have endeavored to explicate how a mechanistic receptor model can be translated into an information theory structure in a way that makes meaningful comparison with experiment possible. More broadly, this project connects to the larger question of whether biological signaling systems are organized according to efficient coding or infomax-like principles. The present results do not show that the NMJ maximizes information transmission. But they do suggest that receptor activation can be studied in those terms, and that the shape of the receptor response curve already places structure on which inputs are most informative. In that sense, we point toward a more extensive view of the NMJ not exclusively as a biochemical system, but as a signaling system whose structure may help determine how information about the input is kept or lost.

\begin{acknowledgments}
We wish to acknowledge Gautam Agarwal for useful comments and encouragement and Mathematica for figure generation \cite{Mathematica}. All Mathematica code is available in the Github MaximalInformationTransmissionNMJ.
\end{acknowledgments}

\appendix

\bibliography{apssamp}

\section{Alterations if equilibrium membrane potential changes}
\label{app:1}

As described in Ref. \cite{bialek2012biophysics}, in theory if statistical mechanics is used, open probabilities of ion channels must change if opening allows current to flow. In general, this can lead to complications, where a binomial distribution no longer describes the conditional probability distribution of the number of open ion channels because the membrane voltage depends in turn on how many channels are open via the cable equation. However, if other ion channels that do not depend on the neurotransmitter act in concert to roughly keep membrane potential the same, then physiological dose-response relationships are adequate for understanding the information channel.

In this appendix, we ask what happens to our predictions if membrane voltage suddenly changes. This is a first step towards addressing how important membrane voltage coupling between ion channels might be. If slight fluctuations in membrane potential greatly change dose-response relationships, then we cannot assume that membrane potential remains at equilibrium and must account for voltage coupling between ion channels.

To understand how much membrane potential matters, we note that if a statistical mechanical model were to incorporate voltage dependence, then the statistical mechanical weight of the open probability becomes biased by a factor of $e^{-QeV/k_B T}$ where $T$ is the temperature, $Q$ is the gating charge, $e$ is the charge of an electron, and $V$ is the membrane voltage. Let's assume that the physiological conditions used in empirically finding dose-response relationships were that $V_{eq}=-85 mV$, $Q=1$, and $T=300 K$. If $V$ were to change, the open statistical mechanical weight would change by a factor of $\frac{e^{-QeV/k_BT}}{e^{-QeV_{eq}/k_BT}} = e^{-Qe(V-V_{eq})/k_B T}$, and the dose-response relationship would change via
\begin{equation}
p_{open}(c) = \frac{ e^{-Qe(V-V_{eq})/k_B T}w_{open}}{e^{-Qe(V-V_{eq})/k_B T}w_{open}+w_{closed}}.
\end{equation}
Here, $w_{open}$ is the statistical mechanical weight of the open state, while $w_{closed}$ is the statistical mechanical weight of the closed state.

Depending on the statistical mechanical model used to capture the dose-response relationship, these open and closed statistical weights will vary. For the Hill model, we find
\begin{equation}
    p_{open}(c) = \frac{e^{-Qe(V-V_{eq})/k_B T} c^{n_H}}{e^{-Qe(V-V_{eq})/k_B T} c^{n_H}+K_d^{n_H}}.
\end{equation}
For a more complicated model, like the Monod-Wyman-Changeux model \cite{marzen2013statistical}, the dose-response relationship would change to
\begin{equation}
    p_{open}(c) = \frac{e^{-Qe(V-V_{eq})/k_B T} \left(1+\frac{c}{K_{d,O}}\right)^2}{e^{-Qe(V-V_{eq})/k_B T} \left(1+\frac{c}{K_{d,O}}\right)^2+e^{-\beta\Delta\epsilon}\left(1+\frac{c}{K_{d,C}}\right)^2}.
\end{equation}
More complicated models that allow for different binding sites can be used as well \cite{prince1999acetylcholine}.

\begin{figure}
\centering
\includegraphics[width=0.5\textwidth]{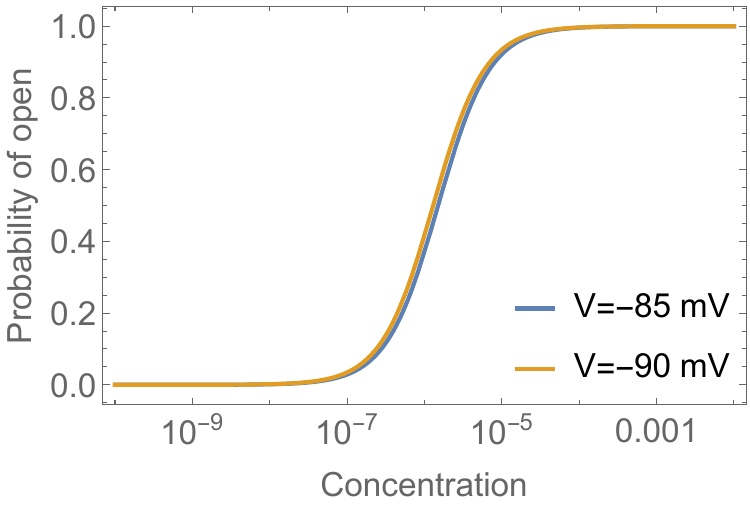}
\includegraphics[width=0.5\textwidth]{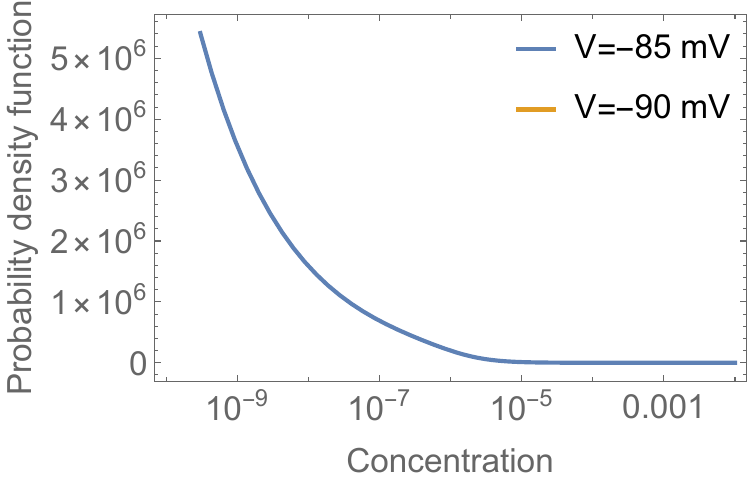}
\caption{Assuming that the dose-response relationship was measured at $-85 mV$, we find that at $-90 mV$ for the membrane potential, the dose-response relationship changes slightly at top but the optimal input distribution hardly changes at bottom. All concentrations in molar.}
\label{fig:fluctuations}
\end{figure}

Optimal input distributions do change when membrane voltage fluctuates, but not by much, as shown in Fig. \ref{fig:fluctuations}. As such, it seems that our approximation to ignore voltage coupling is likely alright for an order-of-magnitude calculation, but this always has to be checked.

\section{Validity of the small-noise approximation}
\label{app:validity}

When using the phenomenological Hill model, a problem with the small-noise approximation arises. Namely, $p_{open}$ dips so low or so close to $1$ that at very low or very high concentrations, the small-noise approximation places a significant chunk of its probability mass outside the physiologically allowed range of $p_{open}\in [0,1]$. To ameliorate this, we augment the phenomenological Hill model so that it is more accurate to dose-response curves by allowing for a dynamic range that is not $1$:
\begin{equation}
p_{open}(c) = p_{open}^{(min)}+(p_{open}^{(max)}-p_{open}^{(min)}) \frac{c^{n_H}}{c^{n_H}+K_d^{n_H}}.
\label{eq:Hill2}
\end{equation}
This new phenomenological model does not alter any subsequent part of the calculation except to make the small-noise approximation valid for all concentrations $c$ in the biophysically justified circumstance that the dynamic range is high but not $1$. For instance, the dose-response curve of nACHRs can also be modeled by the Monod-Wyman-Changeux model, which necessarily has a dynamic range less than $1$ \cite{marzen2013statistical}.
In the experiments that found dose-response curves for these neurotransmitters, the dynamic range was near $1$, and so the offset to the channel capacity approaches its maximal value of $\log_2 \sqrt{\frac{\pi}{2e}}$. The calculations in the main text can thus be viewed as having implicitly used the phenomenological model in Eq. \ref{eq:Hill2}, then taking the limit as the dynamic range approaches $1$ of the channel capacity and optimal input distribution.

\end{document}